\documentclass[aps,preprintnumbers,nofootinbib,superscriptaddress,floatfix]{revtex4}

\usepackage{amssymb}
\usepackage{amsmath}
\usepackage{bm}
\usepackage{slashed}
\usepackage{datetime}
\usepackage{mciteplus}
\usepackage{multirow}
\usepackage{color}
\usepackage[dvipsnames]{xcolor}
\usepackage[utf8]{inputenc}
\usepackage[normalem]{ulem}
\usepackage{graphicx}
\usepackage{hyperref}

\pdfoutput=1 
             
\begin{document}
\allowdisplaybreaks[2]

\title{A first extraction of gluon TMDs from Higgs data at the LHC \\ \vspace{0.2cm}
\normalsize{\textmd{The \textbf{MAP} (Multi-dimensional Analysis of Partonic distributions) Collaboration}}}

\author{Simone Anedda}
\thanks{Electronic address: sanedda@dsf.unica.it --\href{https://orcid.org/0009-0005-4921-2850}{ORCID: 0009-0005-4921-2850}}
\affiliation{Dipartimento di Fisica, Universit\`a di Cagliari, Cittadella Universitaria, I-09042 Monserrato (CA), Italy}
\affiliation{INFN - Sezione di Cagliari, Cittadella Universitaria, I-09042 Monserrato (CA), Italy}

\author{Valerio Bertone}
\thanks{Electronic address: valerio.bertone@cea.fr -- \href{https://orcid.org/0000-0003-0148-0272}{ORCID: 0000-0003-0148-0272}}
\affiliation{Université Paris-Saclay - CEA - IRFU, 91191 Gif-sur-Yvette, France}

\author{Giuseppe Bozzi}
\thanks{Electronic address: giuseppe.bozzi@unica.it -- \href{https://orcid.org/0000-0002-2908-6077}{ORCID: 0000-0002-2908-6077}}
\affiliation{Dipartimento di Fisica, Universit\`a di Cagliari, Cittadella Universitaria, I-09042 Monserrato (CA), Italy}
\affiliation{INFN - Sezione di Cagliari, Cittadella Universitaria, I-09042 Monserrato (CA), Italy}

\author{Matteo Cerutti}
\thanks{Electronic address: matteo.cerutti@cea.fr --\href{https://orcid.org/0000-0001-7238-5657}{ORCID: 0000-0001-7238-5657}}
\affiliation{Université Paris-Saclay - CEA - IRFU, 91191 Gif-sur-Yvette, France}

\begin{abstract}
  We present the first extraction of the unpolarised gluon
  transverse-momentum-dependent (TMD) parton distribution from
  Higgs-boson production data at the LHC within the framework of TMD
  factorisation. The analysis is based on the currently available set of ATLAS and CMS
  measurements of the Higgs $q_T$ distribution at $\sqrt{s} = 8$ and
  $13$ TeV in the diphoton and four-lepton decay channels, restricted
  to the small-$q_T$ region where TMD factorisation is
  applicable. Theoretical predictions are computed up to N$^3$LL
  accuracy, with the contribution of the linearly polarised gluon TMD
  $h_1^{\perp g}$ accounted for. Fiducial selections are consistently
  incorporated for both two- (diphoton) and four-body (four-lepton)
  final states. The fit reproduces both the shape and the
  normalisation of the experimental data, and yields a moderate
  sensitivity to the nonperturbative content of gluon TMDs. We further
  assess the convergence of the perturbative expansion and the
  stability of the extracted distribution under variations of the
  $q_T$ cut. This analysis provides a baseline for future extractions
  combining LHC Higgs measurements with other gluon-sensitive
  processes spanning a broader range of hard scales.
\end{abstract}

\maketitle

\section{Introduction}

The internal structure of the proton is described, at leading power in the hard scale, by parton distribution functions (PDFs). When the transverse momentum of partons with respect to the proton direction of motion is resolved, this description is extended to transverse-momentum dependent distributions (TMDs), which encode both the longitudinal momentum fraction $x$ and the intrinsic transverse momentum $\boldsymbol{k}_\perp$ of partons inside a fast-moving hadron. TMD factorisation theorems~\cite{Collins:1984kg,Collins:2011zzd} allow us to express the differential cross section for suitable processes characterised by an inclusive final state with invariant mass $Q$ and transverse momentum $q_T \equiv |\mathbf{q}_T| \ll Q$ as convolutions of TMD distributions with perturbatively calculable hard coefficients. Importantly, TMDs obey specific evolution equations whose solution gives rise to the characteristic resummation of large logarithms $\ln(Q^2/q_T^2)$.

Whereas quark TMDs have been extensively studied and extracted from Drell-Yan and semi-inclusive deep-inelastic scattering data over the past two decades~\cite{Bacchetta:2017gcc,Bertone:2019nxa,Scimemi:2019cmh,Bacchetta:2019sam,Bacchetta:2022awv,Bury:2022czx,Moos:2023yfa,Bacchetta:2024qre,Moos:2025sal,Bacchetta:2025ara}, gluon TMDs remain far less constrained. This asymmetry can mostly be ascribed to the scarcity of experimental information. As a matter of fact, a clean extraction of the unpolarised gluon TMDs requires processes whose final state is colour-singlet, so that soft and collinear radiation factorises without ambiguity~\cite{Catani:1999ss,Catani:2000vq,Bozzi:2005wk}. Conversely, colour-entanglement issues can compromise TMD factorisation in processes with coloured final states such as jets or quarkonia. Nonetheless, a broad literature exists which includes studies of the gluon TMDs such as quarkonium production at the LHC~\cite{Scarpa:2019fol, Lansberg:2017dzg}, gluon linear-polarisation effects~\cite{Sun:2011iw, Boer:2011kf, Boer:2013fca, Boer:2014lka, Boer:2014tka}, and model calculations~\cite{Bacchetta:2020vty,Bacchetta:2024fci}.

Inclusive Higgs production in proton-proton collisions is possibly the cleanest probe of the unpolarised gluon TMDs. Indeed, at the LHC the Higgs boson is predominantly produced through gluon fusion. Taking into account only electroweak decay modes, the observed colour-singlet final state avoids colour-entanglement issues that could hamper TMD factorisation.

The theoretical description of the Higgs $q_T$ distribution has a long history. At small $q_T$, the resummation of large logarithms of $q_T$ is necessary to obtain reliable perturbative predictions. The groundwork for such resummation in the $b$-space formalism was laid by Collins, Soper and Sterman (CSS)~\cite{Collins:1984kg}, and the Higgs case was first worked out explicitly at next-to-leading logarithmic (NLL) accuracy in Refs.~\cite{Kauffman:1991jt,Yuan:1991we}. A comprehensive framework combining resummation at small $q_T$ with fixed-order matching at large $q_T$ was developed in Ref.~\cite{Bozzi:2005wk}, who computed the Higgs $q_T$ distribution at next-to-next-to-leading logarithmic (NNLL) accuracy matched to next-to-leading order (NLO). Subsequent advances pushed the accuracy to N$^3$LL+NNLO, achieved with different methods, such as momentum-space resummation in Ref.~\cite{Bizon:2018foh}, and soft-collinear effective theory (SCET) in Ref.~\cite{Billis:2021ecs}. These developments established the theoretical infrastructure required for precision phenomenology of the Higgs $q_T$ distribution.

In the small-$q_T$ region, the differential Higgs cross section becomes directly sensitive to the unpolarised gluon TMD $f_1^g$ and to the linearly polarised gluon TMD $h_{1}^{\perp g}$ (Boer-Mulders function). The linearly polarised gluon TMD was first studied in Ref.~\cite{Mulders:2000sh}, where it was shown that an unpolarised hadron can contain gluons with linear polarisation in the transverse plane, giving rise to an independent TMD that contributes to the cross section on equal footing with $f_1^g$. The matching of $h_1^{\perp g}$ onto collinear gluon PDFs was computed at NLO in Refs.~\cite{Catani:2010pd,Becher:2012yn} and pushed to NNLO in Ref.~\cite{Gutierrez-Reyes:2019rug}, completing the perturbative ingredients needed for a consistent NNLO treatment of both gluon TMDs in Higgs production.

Despite this theoretical progress, a direct extraction of gluon TMDs from experimental data remains an open challenge. The reason is partly experimental: precision measurements of the Higgs $q_T$ distribution have only recently become available from the ATLAS and CMS experiments at Run I and Run II of the LHC, covering both the di-photon and four-lepton decay channels. These datasets, while limited in statistical power, now constitute a viable basis for a first phenomenological extraction.

In this paper, we present the first extraction of the unpolarised gluon TMD $f_1^g$ from LHC Higgs data within the TMD factorisation framework, employing perturbative accuracy up to N$^3$LL. We consider the complete set of available ATLAS and CMS measurements of the Higgs $q_T$ distribution in the $H\to\gamma\gamma$ and $H\to 4\ell$ decay channels, and perform a fit using the NangaParbat framework~\cite{Bacchetta:2022awv} developed by the MAP Collaboration. We assess the quality of the fit, the nonperturbative sensitivity of the current dataset, and the convergence of the perturbative series from NLL' to N$^3$LL. The linearly polarised gluon contribution is included in the theoretical predictions with its NNLO matching coefficient, and its numerical impact on the extracted distributions is discussed.

The paper is organised as follows: section~II provides an overview of the TMD formalism used in the analysis; section~III describes the datasets considered in the fitting procedure and discusses the implementation of the selection cuts in our theoretical framework; section~IV presents the numerical results of the fit and the $k_T$ shape of the extracted gluon TMD; in section~V we draw our conclusions and elaborate on possible extensions of the present work.

\section{Formalism}
\label{s:formalism}

We consider the following process
\begin{equation}
  p(P_1)\,+\,p(P_2)\to H(q)\,+\,X\,,
\end{equation}
in which two protons with four-momenta $P_{1}$ and $P_2$, such that $(P_1+P_2)^2=s$, collide to inclusively produce an on-shell Higgs boson of invariant mass $\sqrt{q^2}=M_H=125.1$~GeV, rapidity $y$, and transverse momentum $\mathbf{q}_T$.

\begin{figure}[h]
    \centering
    \includegraphics[scale=0.6]{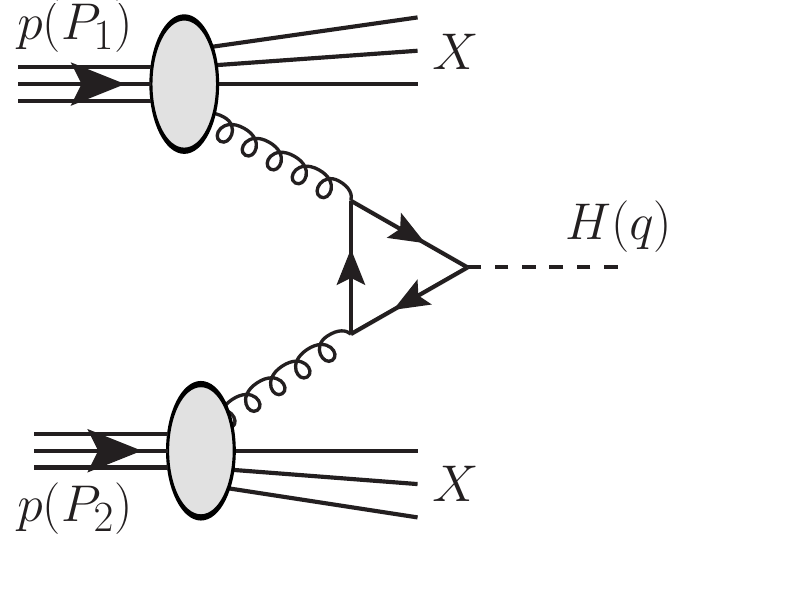}
    \caption{Higgs production in $pp$ collisions through gluon fusion at leading order.}
    \label{fig:gluon fusion}
\end{figure}

The dominant partonic production channel at the LHC is gluon fusion, $gg\to H$, mediated at leading order by a heavy quark loop, as illustrated in Fig.~\ref{fig:gluon fusion}. Since the Yukawa coupling of the Higgs boson to fermions is proportional to the fermion mass, the amplitude is dominated by a virtual top loop, and contributions from lighter quarks (e.g. $b,c$) can be neglected. 

The LO matrix element squared for Higgs production in gluon fusion is~\cite{Ellis:1996mzs}
\begin{equation}
  |M(gg\to H)|^2=\frac{\alpha_S^2(M_H)M_H^4G_F}{288\pi\sqrt{2}}
  \left|A_Q\!\left(\frac{4M_t^2}{M_H^2}\right)\right|^{2},
\end{equation}
where $\alpha_{s}(M_{H})$ is the strong coupling evaluated at the Higgs mass, $G_F=1.16639\times 10^{-5}\,\mathrm{GeV}^{-2}$ is the Fermi constant, and $M_t=172.9$~GeV denotes the on-shell pole mass of the top. The function $A_Q$ is given by
\begin{equation}
    \begin{aligned}
    A_Q(x)&=\frac{3}{2}\,x\Bigl[1+(1-x)f(x)\Bigr],\\
        f(x)&=
\begin{cases}
\arcsin^{2}\!\left(\dfrac{1}{\sqrt{x}}\right), & x\ge 1,\\[6pt]
-\dfrac{1}{4}\left[\ln\!\left(\dfrac{1+\sqrt{1-x}}{1-\sqrt{1-x}}\right)-i\pi\right]^{2}, & x<1.
\end{cases}
\end{aligned}
\label{eq:fdecayfunction}
\end{equation}

In the small transverse momentum region, $q_T\ll M_H$ with $q_T=|\mathbf{q}_T|$, the differential cross section for the production of an Higgs boson can be factorised in terms of TMDs. Moreover, we use the narrow-width approximation (NWA) which sets the Higgs on the mass shell, producing
\begin{equation}
  \begin{array}{rcl}
    \displaystyle\frac{d\sigma}{dydq_T^2}&=&\displaystyle \mathcal{P}(M_H,y,q_T)\frac{\alpha_S^2(M_H)G_F}{288\pi\sqrt{2}}H_{ggH}(M_H,\mu)\,\int_0^\infty db_T\,b_TJ_0(q_Tb_T)\\
    \\
    \displaystyle&\times&\displaystyle\left[f_1^g(x_1,b_T;\mu,\zeta_1)\,f_1^g(x_2,b_T;\mu,\zeta_2)\, + {h}_1^{\perp g}(x_1, b_T;\mu,\zeta_1)\,{h}_1^{\perp g} (x_2, b_T;\mu,\zeta_2)\right]\,,
  \end{array}
  \label{eq:higgsfactoriz}
\end{equation}
where $x_{1,2}=(M_H/\sqrt{s})e^{\pm y}$ are the light-cone momentum fractions. $\mathcal{P}$ is the phase-space reduction factor associated with the kinematic cuts on the final state which will be discussed in section~\ref{s:data}. The perturbative hard function $H_{ggH}$ embodies hard and virtual contributions to the cross section and its perturbative coefficients up to $\mathcal{O}(\alpha_s^2)$ can be found in Ref.~\cite{Bizon:2018foh}. The unpolarised and linearly polarised (Boer-Mulders) TMDs ${f}_1^g$ and ${h}_1^{\perp g}$, respectively, are expressed in $\mathbf{b}_T$ space (with $b_T \equiv |\mathbf{b}_T|$) Fourier-conjugate to $\mathbf{q}_T$. Finally, the renormalisation scale $\mu$ and the two rapidity scales $\zeta_1$ and $\zeta_2$ are set as follows: $\mu^{2}=\zeta_1=\zeta_2=M_H^2$.

As is well known, for small values of $b_T$ both TMDs ${f}_1^g$ and ${h}_1^{\perp g}$ can be matched on the unpolarised collinear PDFs. Specifically, one has:
\begin{equation}
  \begin{array}{l}
    \displaystyle f_1^g(x,b_T;\mu_b,\mu_b^2)=\sum_{i=g,q}\int_x^1\frac{dy}{y}C_{gi}(y,\alpha_s(\mu_b))f_i\left(\frac{x}{y},\mu_b\right)\equiv [C\otimes f]_g(x,b_T;\mu_b,\mu_b^2)\,,\\
    \\
    \displaystyle h_1^{\perp g}(x,b_T;\mu_b,\mu_b^2)=\sum_{i=g,q}\int_x^1\frac{dy}{y}G_{gi}(y,\alpha_s(\mu_b))f_i\left(\frac{x}{y},\mu_b\right) \equiv [G\otimes f]_g(x,b_T;\mu_b,\mu_b^2)\,,
  \end{array}
  \label{eq:TMDmatching}
\end{equation}
where $\mu_b=2e^{-\gamma_{\rm E}}/b_T$, with $\gamma_{\rm E}$ the Euler-Mascheroni constant, and $f_i$ correspond to the unpolarised collinear PDFs, which are taken from the NNPDF3.1 set~\cite{NNPDF:2017mvq} with perturbative charm at the appropriate perturbative order~\cite{Bacchetta:2019sam} and accessed through the {\tt LHAPDF} library~\cite{Buckley:2014ana}. Also the running of the strong coupling $\alpha_s$ is taken from the PDF set. The matching functions $C_{gi}$ and $G_{gi}$ admit the following perturbative expansions:
\begin{equation}
  \begin{array}{l}
    \displaystyle C_{gi}(y,\alpha_s)=\delta(1-y)+\sum_{n=1}^\infty\left(\frac{\alpha_s}{4\pi}\right)^nC_{gi}^{[n]}(y)\,,\\
    \\
    \displaystyle G_{gi}(y,\alpha_s) =\sum_{n=1}^\infty\left(\frac{\alpha_s}{4\pi}\right)^nG_{gi}^{[n]}(y)\,,
  \end{array}
\end{equation}
which we truncate at $\mathcal{O}(\alpha_s)$ for NLL' and NNLL, and at $\mathcal{O}(\alpha_s^2)$ for NNLL' and N$^3$LL. Moreover, TMDs obey evolution equations that govern their dependence on $\mu$ and $\zeta$. The solution to these evolution equations is encoded in the so-called Sudakov form factor, which is the same for both unpolarised and linearly-polarised gluon TMDs and reads
\begin{equation}
R^g\left[\left(\mu_b,\mu_b^2\right)\rightarrow(\mu,\zeta)\right]=\exp\left\{K^g(\mu_b)\ln\frac{\sqrt{\zeta}}{\mu_b}+\int_{\mu_b}^{\mu}\frac{d\mu'}{\mu'}\left[\gamma_F^g(\alpha_s(\mu'))-\gamma_K^g(\alpha_s(\mu'))\ln\frac{\sqrt{\zeta}}{\mu'}\right]\right\}\,.
\end{equation}
The gluon anomalous dimensions $K^g$ (Collins-Soper), $\gamma_F^g$ (non-cusp), and $\gamma_K^g$ (cusp) also admit a perturbative expansion and are truncated at the appropriate order depending on whether the cross-section computation is carried out at NLL', NNLL, NNLL', or N$^3$LL~\cite{Bacchetta:2019sam}.

Finally, TMDs can be computed as follows:
\begin{equation}
  \begin{array}{l}
    \displaystyle f_1^g(x,b_T;\mu,\zeta)=R^g\left[\left(\mu_b,\mu_b^2\right)\rightarrow(\mu,\zeta)\right][C\otimes f]_g(x,b_T;\mu_b,\mu_b^2)\,,\\
    \\
    \displaystyle h_1^{\perp g}(x,b_T;\mu,\zeta)=R^g\left[\left(\mu_b,\mu_b^2\right)\rightarrow(\mu,\zeta)\right][G\otimes f]_g(x,b_T;\mu_b,\mu_b^2)\,.
  \end{array}
\end{equation}
However, when $b_T$ becomes large, or equivalently when $\mu_b$ becomes small, these expressions become unreliable because of the onset of nonperturbative effects. Indeed, in this region the strong coupling approaches the Landau pole and the matching formulas in Eq.~(\ref{eq:TMDmatching}) receive large power corrections. In order to account for nonperturbative effects, we follow the traditional CSS strategy~\cite{Collins:1984kg} by introducing the function~\cite{Bacchetta:2017gcc}
\begin{equation}
 b_*(b_T)= b_{\rm max} \left( \frac{1 - e^{-b_T^4 / b_{\rm max}^4}}{1 - e^{-b_T^4 / b_{\rm min}^4}} \right)^{1/4}\,,
\end{equation}
with $b_{\rm max}=2e^{-\gamma_{\rm E}}$ and $b_{\rm min}=2e^{-\gamma_{\rm E}}/M_H$, and rewriting the TMDs as follows:
\begin{equation}
  \begin{array}{l}
    \displaystyle f_1^g(x,b_T;\mu,\zeta)\to f_1^g(x,b_*(b_T);\mu,\zeta) f_{\rm NP}(x,b_T,\zeta) \,,\\
    \\
    \displaystyle h_1^{\perp g}(x,b_T;\mu,\zeta)\to h_1^{\perp g}(x,b_*(b_T);\mu,\zeta) f_{\rm NP}(x,b_T,\zeta)\,.
  \end{array}
\end{equation}
Since $b_*(b_T)$ never becomes too large even for large values of $b_T$, the replacements above guarantee that TMDs computed in $b_*(b_T)$ are always in a regime where perturbative matching and evolution are accurate. Nonperturbative effects are instead factored out into the function $f_{\rm NP}$, which needs to be parametrised and fitted to data (for more details see, \textit{e.g.}, Refs.~\cite{Bacchetta:2019sam,Cerutti:2026apy}). In this study, we choose a simple Gaussian parametrisation:
\begin{equation}
  f_{\rm NP}(x,b_T,\zeta)=\exp\left[-\frac{1}{2}gb_T^2\right] \, ,
  \label{e:fNP_oneparam}
\end{equation}
with $g$ a free parameter to be determined from data. Two observations are in order. The first is that $f_{\rm NP}$ is assumed to be the same for both unpolarised and linearly-polarised TMDs. In general, this might not be the case. However, current data does not allow to determine $f_{\rm NP}$ separately for the two TMDs involved in the calculation. The second observation is that, in general, $f_{\rm NP}$ is also a function of the rapidity scale $\zeta$. In fact, the $\zeta$ dependence of $f_{\rm NP}$ determines the non perturbative contribution to the Collins-Soper kernel (see, \textit{e.g.}, Ref.~\cite{Bacchetta:2025ara}). However, within the NWA, data are considered at one single scale, \textit{i.e.} $\zeta=M_H^2$. As a consequence, we have no way to determine the $\zeta$ dependence of $f_{\rm NP}$. Therefore, Eq.~(\ref{e:fNP_oneparam}) is to be considered valid only at $\zeta=M_H^2$.  

\section{Experimental Data}
\label{s:data}

In this section we describe the experimental datasets included in our
analysis.  We consider measurements of the Higgs-boson
transverse-momentum ($q_T$) distribution performed by the ATLAS and
CMS experiments in the diphoton ($\gamma\gamma$) and four-lepton
($4\ell$) decay channels. The breakdown of the full dataset included
in this analysis is summarised in Table~\ref{tab:datasets}.
\begin{table}[h]
\centering
\setlength{\tabcolsep}{4pt}
\renewcommand{\arraystretch}{1.15}
\begin{tabular}{lcccccc}
\hline
Dataset & \multicolumn{3}{c}{$N_{\rm dat}$} & Decay channel & $\sqrt{s}$ [GeV] & Ref. \\
 & $q_T/M_H<0.2$ & $q_T/M_H<0.25$ & $q_T/M_H<0.3$ & & & \\
\hline
CMS Run II (2017--2018)   & 5 & 6 & 7 & $H\to \gamma\gamma$ & 13000 & \cite{CMS:2022wpo} \\
CMS Run II (2017--2018)   & 2 & 3 & 3 & $H\to 4\ell$ & 13000 & \cite{CMS:2023gjz} \\
CMS Run II (2015--2016)   & 1 & 2 & 2 & combined & 13000 & \cite{CMS:2018gwt} \\
CMS Run I                 & 1 & 2 & 2 & $H\to \gamma\gamma$ & 800 & \cite{CMS:2015qgt} \\
ATLAS Run II (2017--2018) & 5 & 6 & 7 & $H\to \gamma\gamma$ & 13000 & \cite{ATLAS:2022fnp} \\
ATLAS Run II (2017--2018) & 2 & 3 & 3 & $H\to 4\ell$ & 13000 & \cite{ATLAS:2020wny} \\
ATLAS Run II (2015--2016) & 2 & 3 & 3 & combined & 13000 & \cite{ATLAS:2018pgp} \\
ATLAS Run I               & 1 & 2 & 2 & $H\to \gamma\gamma$ & 800 & \cite{ATLAS:2014yga} \\
ATLAS Run I               & 1 & 1 & 1 & $H\to 4\ell$ & 800 & \cite{ATLAS:2014xzb} \\
\hline
\textbf{Total}            & \textbf{20} & \textbf{28} & \textbf{30} & -- & -- & -- \\
\hline
\end{tabular}
\caption{Breakdown of the datasets included in this analysis. For each
  dataset, the table includes information on: the number of data
  points ($N_{\mathrm{dat}}$) passing three different cuts in
  $q_T/M_H$, the Higgs-boson decay channel, the center-of-mass energy
  $\sqrt{s}$, and the reference.}
\label{tab:datasets}
\end{table}

The datasets considered span different center-of-mass energies
($\sqrt{s}=8$ and $13$~TeV) and probe phase-space regions
characterised by an invariant mass of the Higgs boson around its
on-shell mass $M_H$ and typical momentum fractions $x \sim 10^{-2}$.
As for the ATLAS measurements, we include only combined results when
available, in order to avoid double counting.  Since the analysis is
performed within the NWA, the experimental data provides information
only at a single hard scale, $M_H$.  Also, the included datasets are
integrated in rapidity $y$ and thus provide very little sensitivity to
the $x$ dependence of TMDs.

Our analysis is restricted to the region of validity of TMD
factorisation (Eq.~\eqref{eq:higgsfactoriz}). As a consequence, we
need to require an upper value on the value of $q_T$ of the data
points included in the analysis. As a baseline, we choose to cut all
data points which do not respect the cut $q_T/M_H < 0.3$. However, we
also consider two more conservative cuts, namely $q_T/M_H < 0.25$ and
$q_T/M_H < 0.2$. Since measurements are provided in finite
transverse-momentum bins, the cut is conservatively applied using the
upper edge of each $q_T$ bin.  The numbers of data points satisfying
the $q_T$-cut conditions for all of the choices discussed above are
reported in Table~\ref{tab:datasets} for the single datasets and for
the full ensemble. With the baseline cut ($q_T/M_H < 0.3$), we include
in the fit a total of 30 points.

All measurements are defined within fiducial phase-space regions,
characterised by kinematic cuts on final-state leptons or
photons. These effects are accounted for in the theory predictions
through the phase-space reduction factor $\mathcal{P}$ (see
Eq.~\eqref{eq:higgsfactoriz} and Appendix~\ref{sec:PSRED}). Combined
measurements, which are extrapolated to the full phase space, are
treated accordingly without fiducial cuts.
Tables~\ref{tab:leptoncuts}-\ref{tab:photoncuts} summarise the
fiducial cuts implemented for each channel and experiment.

\begin{table}[h]
\centering
\setlength{\tabcolsep}{5pt}
\renewcommand{\arraystretch}{1.15}
\begin{tabular}{lcccc}
\hline
\textbf{Dataset} & $p_{T,\ell}$ [GeV] & $|\eta_\ell|$ & Event $p_T$ [GeV] & $m_{\ell\ell}$ [GeV] \\
\hline
CMS Run II (2017--2018) 
& $>7\,(e),\;>5\,(\mu)$ 
& $<2.5\,(e),\;\;\,<2.4\,(\mu)$ 
& $>20,\;>10$ 
& $40<m_{12}<120,\;12<m_{34}<120$ \\

ATLAS Run I 
& $>7\,(e),\;>6\,(\mu)$ 
& $<2.47\,(e),\;<2.7\,(\mu)$ 
& $>20,\;>15,\;>10$ 
& $50<m_{12}<106,\;12<m_{34}<115$ \\

ATLAS Run II (2017--2018) 
& $>7\,(e),\;>5\,(\mu)$ 
& $<2.47\,(e),\;<2.7\,(\mu)$ 
& $>20,\;>15,\;>10$ 
& $50<m_{12}<106,\;12<m_{34}<115$ \\
\hline
\end{tabular}
\caption{Fiducial selection requirements for the $H\to 4\ell$ channel.}
\label{tab:leptoncuts}
\end{table}

\begin{table}[h!]
\centering
\setlength{\tabcolsep}{5pt}
\renewcommand{\arraystretch}{1.15}
\begin{tabular}{lccr@{$\;\lor\;$}l}
\hline
\textbf{Dataset} & $p_{T,\gamma_1}/m_{\gamma\gamma}$ & $p_{T,\gamma_2}/m_{\gamma\gamma}$ & \multicolumn{2}{c}{$|\eta_\gamma|$} \\
\hline
CMS Run II (2017--2018) 
& $>0.35$ & $>0.25$ 
& $|\eta_\gamma|<1.4442$ & $1.566<|\eta_\gamma|<2.5$ \\

CMS Run I 
& $>0.35$ & $>0.25$ 
& $|\eta_\gamma|<1.44$ & $1.57<|\eta_\gamma|<2.5$ \\

ATLAS Run I 
& $>0.35$ & $>0.25$ 
& $|\eta_\gamma|<1.37$ & $1.56<|\eta_\gamma|<2.37$ \\

ATLAS Run II (2017--2018) 
& $>0.35$ & $>0.25$ 
& $|\eta_\gamma|<1.37$ & $1.52<|\eta_\gamma|<2.37$ \\
\hline
\end{tabular}
\caption{Fiducial selection requirements for the $H\to \gamma\gamma$ channel.}
\label{tab:photoncuts}
\end{table}

Experimental uncertainties include both uncorrelated and correlated
components which are fully incorporated in our analysis following the
treatment adopted in past TMD analyses, (\textit{e.g}, see
Refs.~\cite{Bacchetta:2019sam,Bacchetta:2022awv,Cerutti:2022lmb,Bacchetta:2024qre,Bacchetta:2024yzl,Bacchetta:2025ara,Rossi:2025pwh,Camarda:2025lbt}).
The collinear PDF set used in the computation of TMDs
(NNPDF3.1~\cite{NNPDF:2017mvq}, in our case) carries additional
uncertainties, which in principle should be accounted for. However, we
found that PDF uncertainties are at least one order of magnitude
smaller than the experimental ones. Therefore, we decided not to
include them.

We finally stress that no other normalisation factors have been
applied in this analysis, with the consequence that both shape and
normalisation of the experimental measurements have an impact on the
fit.

\section{Results}
\label{s:results}

In this section, we present the results of the extraction of unpolarised gluon TMDs from the current available measurements of Higgs production (see Sec.~\ref{s:data}). In Sec.~\ref{ss:fitquality}, we discuss the quality of the fit at N$^3$LL in terms of theory/data agreement. In Sec.~\ref{ss:gluonTMDs}, we present the extracted gluon TMDs from our analysis. In Sec.~\ref{ss:pertconv}, we discuss the convergence of perturbative corrections. Finally, in Sec.~\ref{ss:qTcut}, we study the dependence of the fit results on the $q_T$ cut.

\subsection{Fit quality}
\label{ss:fitquality}

In this section, we present the quality of the baseline fit at N$^3$LL with cut $q_T / M_H < 0.3$. The error analysis makes use of the bootstrap method, namely by fitting an ensemble of $N_{\rm rep}=200$ of Monte Carlo replicas of the experimental data. Although the most complete statistical information is represented by the full ensemble of replicas, we choose the $\chi^2$ of the central replica, representing the best fit to the unfluctuated data, as the best estimator of fit quality.

Tab.~\ref{tab:chi2} shows the breakdown of $\chi^2$'s normalised to the number $N_{\rm dat}$ of data points for each of the included measurements and for the full dataset. We note that the value of the global $\chi^2/N_{\rm dat} = 1.49$, corresponding to a $p$-value of about $3\%$, indicates that the fit is able to reasonably reproduce the overall shape and normalisation of the datasets. Some residual tension is observed, which can be attributed to statistical fluctuations in the region of relatively large $q_T$. This tension can be significantly reduced by adopting a more conservative cut in $q_T/M_H$ (see Sec.~\ref{ss:qTcut} for details).
\begin{table}[h]
\centering
\setlength{\tabcolsep}{4pt}
\renewcommand{\arraystretch}{1.15}
\begin{tabular}{lcc}
\hline
Dataset & $N_{\rm dat}$ & $\chi^2 / N_{\rm dat}$ \\
\hline
CMS Run II $H\to \gamma\gamma$   & 7    & 1.15  \\
CMS Run II $H\to 4\ell$          & 3    & 1.17  \\
CMS Run II (combined)            & 2    & 4.46  \\
CMS Run I $H\to \gamma\gamma$    & 2    & 0.26  \\
ATLAS Run II $H\to \gamma\gamma$ & 7    & 1.52  \\
ATLAS Run II $H\to 4\ell$        & 3    & 0.93  \\
ATLAS Run II (combined)          & 3    & 0.99  \\
ATLAS Run I $H\to \gamma\gamma$  & 2    & 3.58  \\
ATLAS Run I $H\to 4\ell$         & 1    & 0.004 \\
\hline
\textbf{Total}                   & \textbf{30} & 1.49 \\
\hline
\end{tabular}
\caption{Breakdown of the normalised $\chi^2$ to the number of data points $N_{\rm dat}$ of the central replica for each dataset included in the fit.}
\label{tab:chi2}
\end{table}

Concerning the individual experiments, we observe that the description of the $H\to\gamma\gamma$ measurements is generally worse than for the rest of the dataset. In particular, the CMS Run II (combined) and ATLAS Run~I $H\to\gamma\gamma$ data exhibit the largest deviations from the fitted theoretical predictions, and determine a significant increase of the global $\chi^2$.

\begin{figure}[h]
\centering
\includegraphics[width=0.49\linewidth]{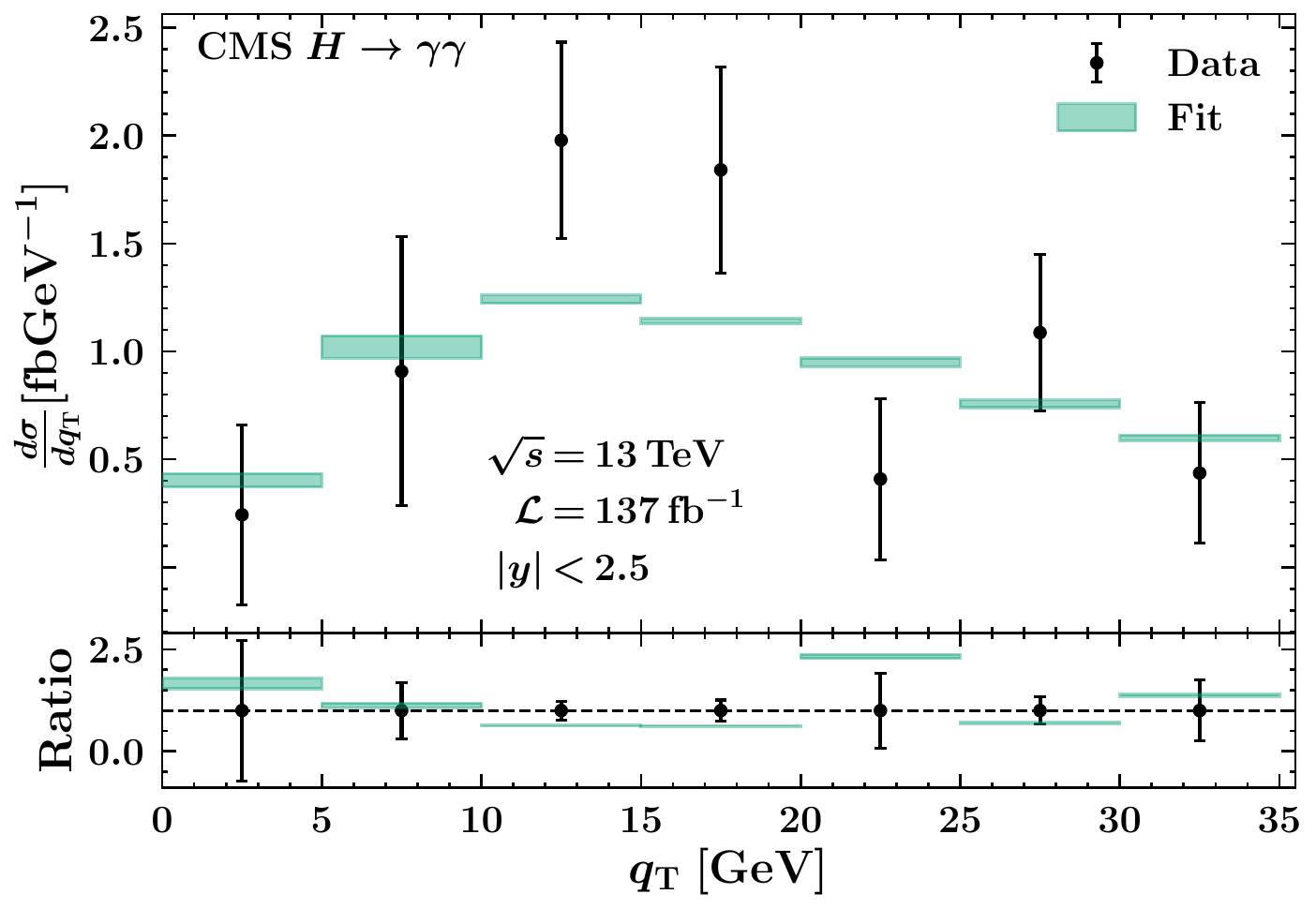}
\includegraphics[width=0.49\linewidth]{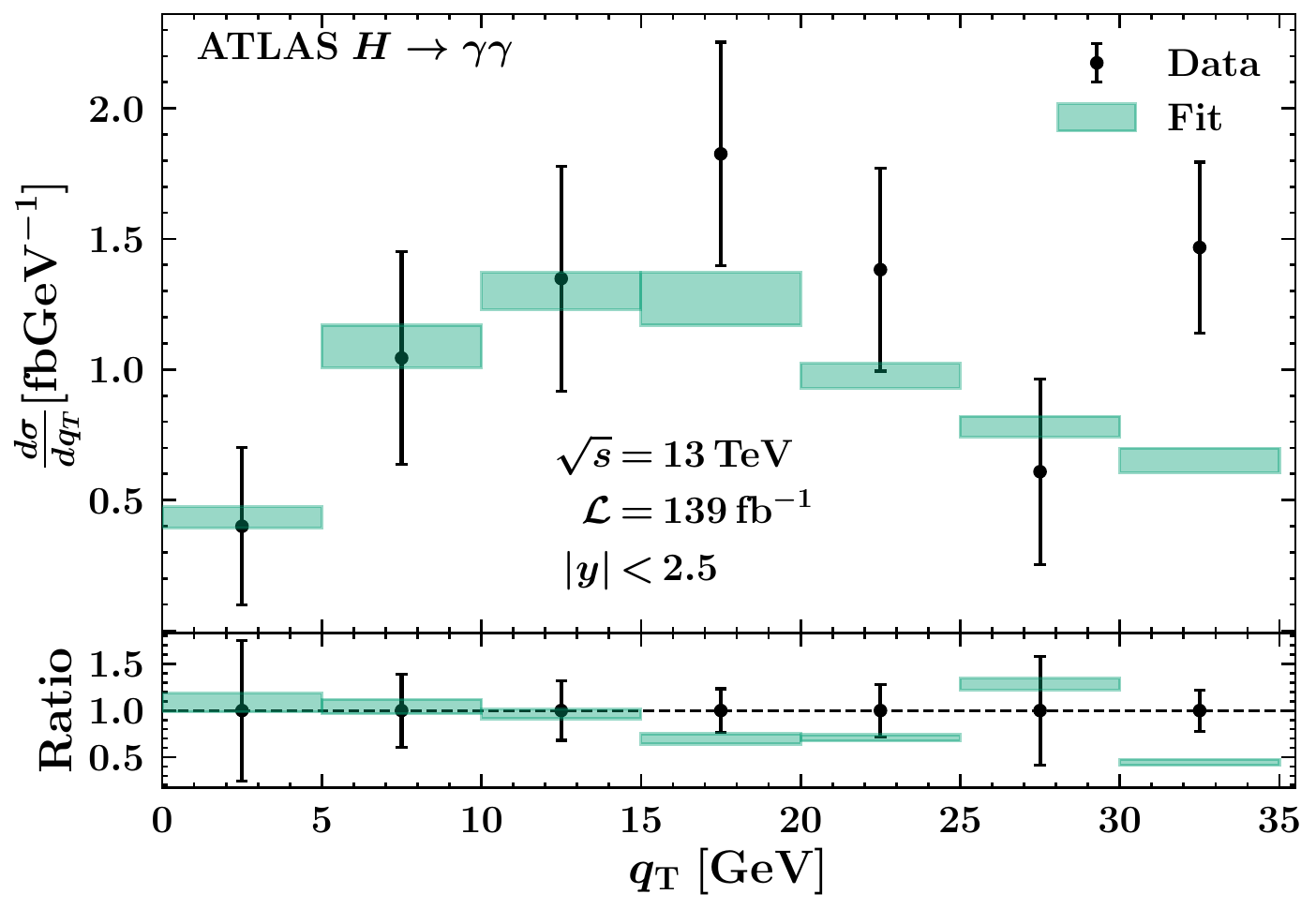}
\caption{Upper panels: comparison between experimental data and theoretical predictions for the $q_T$-differential cross section for Higgs production measurements from CMS Run II (left panel) and ATLAS Run II (right panel);
uncertainty bands correspond to one-sigma error. Lower panel: ratio between experimental data and theoretical results.}
\label{f:DataTheoComp}
\end{figure}

In order to visualize the quality of our fit, we present in Fig.~\ref{f:DataTheoComp} the comparison between experimental data and theoretical results for a representative selection of datasets. We display in the upper panel of each plot the $q_T$-differential cross section, and in the lower panel the data-to-theory ratio.  Coloured bands represent one-sigma uncertainties.

We find that theoretical predictions are able to reproduce both shape and normalisation of data. Some tension is observed in the peak region of CMS Run II $H\to\gamma\gamma$ data and in the tail of the ATLAS Run II $H\to\gamma\gamma$ data. However, as we will prove below, this tension can be significantly reduced by adopting a more conservative $q_T$ cut. Note that the larger theoretical uncertainty bands in the right panel are due to the effect of the systematic shifts induced by correlated experimental errors, which are particularly relevant for the ATLAS Run II $H\to\gamma\gamma$ dataset.

\subsection{Extracted gluon TMDs}
\label{ss:gluonTMDs}

Now we move on to presenting the gluon unpolarised TMD distribution $f_1^g$ extracted from our baseline fit at N$^3$LL with kinematic cut $q_T/M_H<0.3$.\footnote{Also the linearly polarised gluon TMD $h_1^{\perp g}$ is accounted for in this analysis. However, data is little sensitive to this TMD, so we do not find it particularly instructive to discuss it here and draw any conclusions.} In order to gauge the sensitivity of the included experimental data on the intrinsic nonperturbative part of TMDs, we report the value of the free parameter $g$ obtained from the fit:
\begin{equation}
g = 15.6 \pm 5.1 \, \text{GeV}^2 \, ,
\end{equation}
which corresponds to average and standard deviation over the set of 200 Monte Carlo replicas. We observe that the parameter is determined with a relative uncertainty of about $35\%$, which indicates that the current dataset has moderate sensitivity to the nonperturbative part of TMDs. This is consistent with the fact that the included measurements cover a kinematic region at very large scale, where the perturbative contribution to TMDs is dominant. In Sec.~\ref{ss:qTcut}, we will explore the effect on the parameter $g$ of adopting more conservative cuts in $q_T$.

As mentioned above, the NWA prevents the possibility of disentangling the nonperturbative contribution of the Collins-Soper kernel, related to gluon TMD evolution, from the rest of the nonperturbative effects, which we will refer to as intrinsic component. In order to achieve this separation, it is necessary to include measurements spanning different scales broadly different from $M_H$, such as $J/\psi$ production data. We leave this to a future work. Nevertheless, it is instructive to perform an \textit{a posteriori} decomposition, by identifying
\begin{equation}
g = g_1 + g_2^g \log(\zeta/Q_0^2) \, ,
\end{equation}
with $Q_0=1$~GeV and where $g_1$ represents the intrinsic component and $g_2^g$ the nonperturbative evolution. We estimate $g_2^g$ from the quark value extracted in Ref.~\cite{Avkhadiev:2025wps}, $g_2 = 0.167 \pm 0.015~\text{GeV}^2$, rescaled by a factor $C_A/C_F$. This assumption is motivated by perturbative considerations and provides a natural guess in the absence of direct constraints.

We propagate uncertainties via bootstrap sampling of $10^5$ replicas of $(g, g_2^g)$ obtaining
\begin{equation}
g_1 = 14.4 \pm 5.1 \, \text{GeV}^2 \,,
\end{equation}
which allows us to calculate gluon TMDs at values of $\mu=\sqrt{\zeta}=Q$ different from $M_H$. We note that the value of $g_1$ is close to the total $g$, indicating that, within the current kinematic coverage, the nonperturbative evolution contribution is small. This confirms the current limited sensitivity of Higgs measurements on nonperturbative TMD effects.

\begin{figure}[h]
\centering
\includegraphics[width=0.49\linewidth]{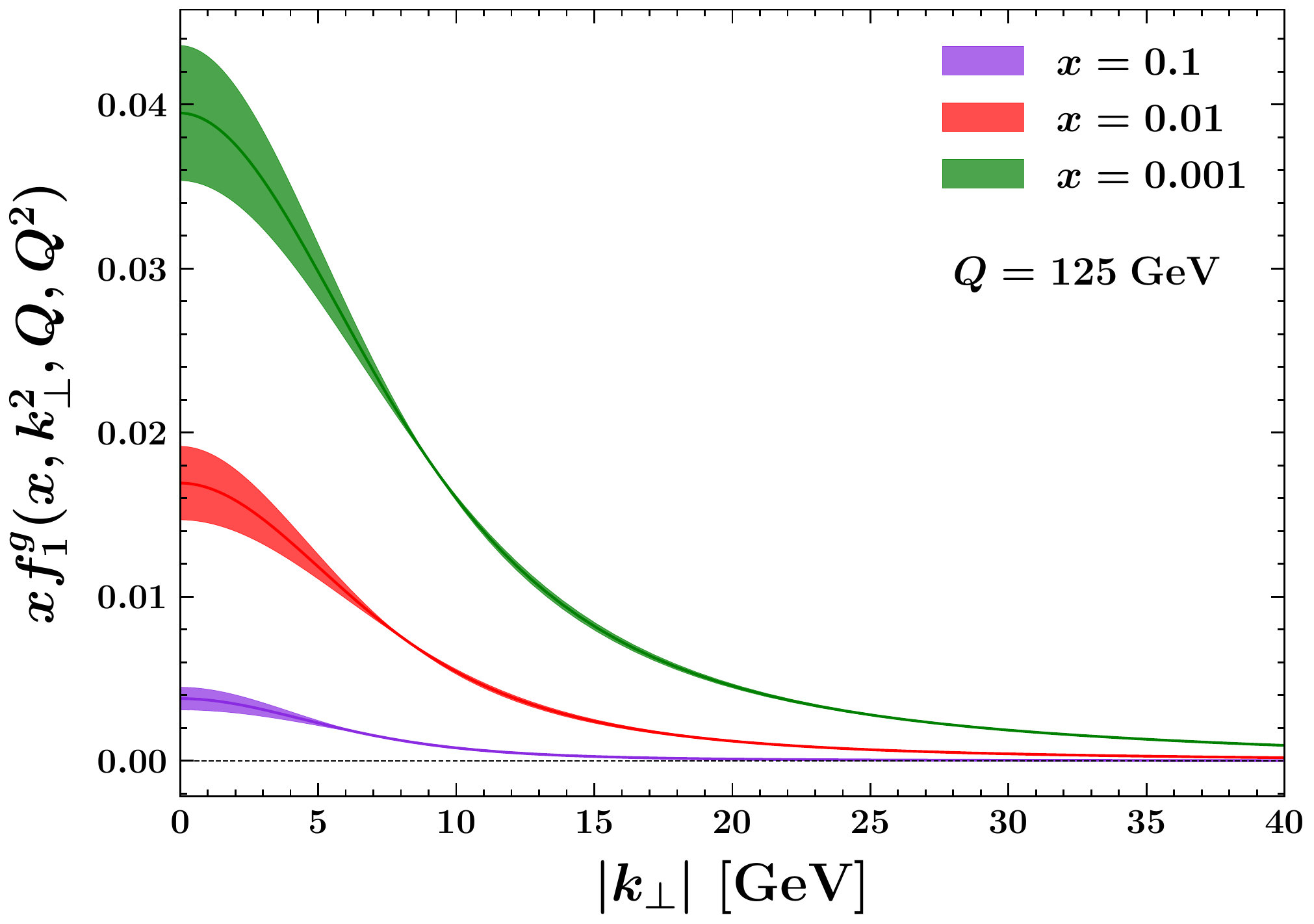}
\includegraphics[width=0.49\linewidth]{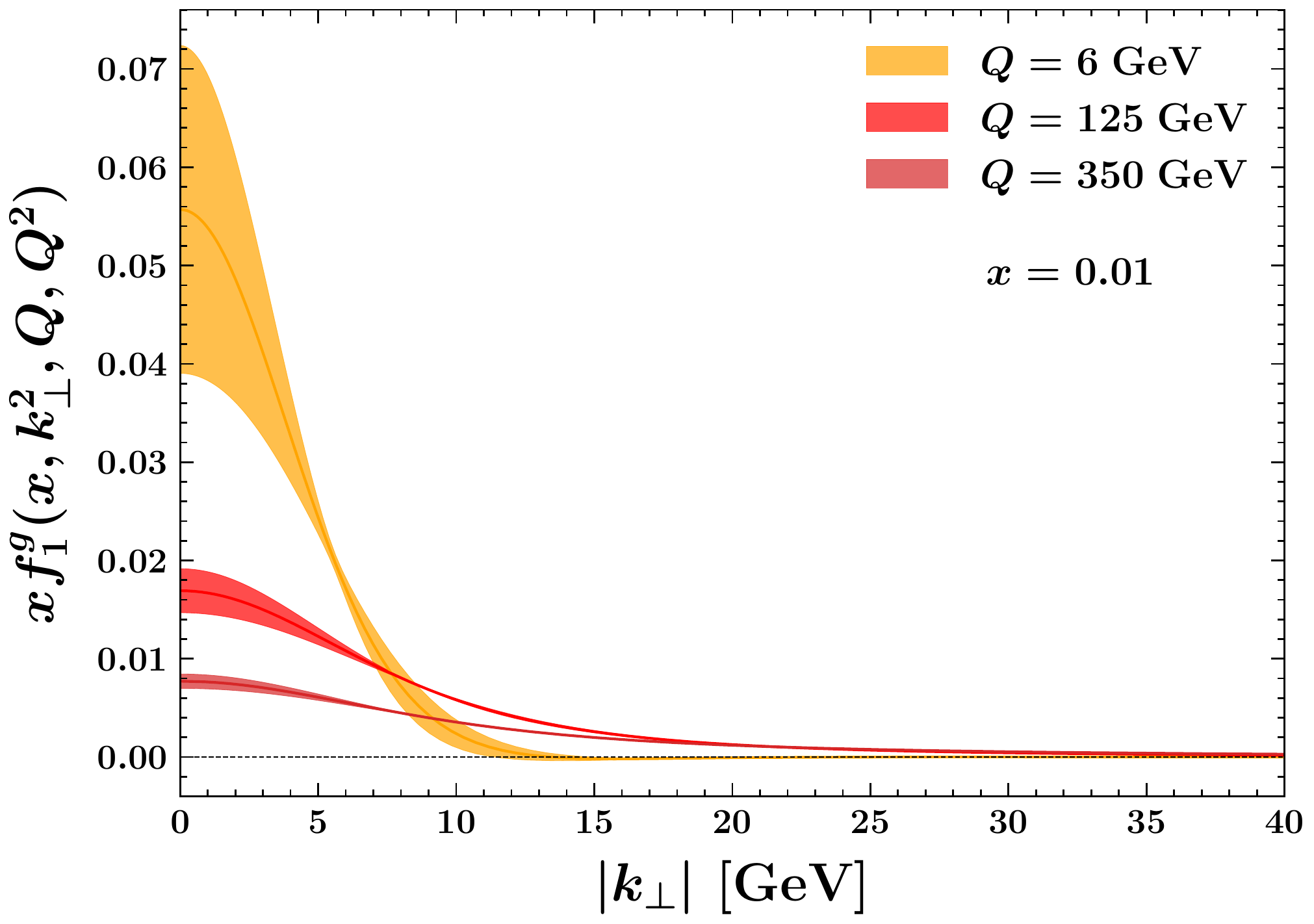}
\caption{Left panel: The TMD PDF of the gluon in a proton at $\mu = \sqrt{\zeta} = 125$ GeV as a function of the partonic transverse momentum $|\boldsymbol{k}_\perp|$ for $x=0.001, 0.01, 0.1$. Right panel: gluon TMD PDF at $x=0.01$ for $\mu = \sqrt{\zeta} = 6, 125, 350$ GeV.
Uncertainty bands correspond to one-sigma error.}
\label{f:gluonTMDs}
\end{figure}

In Fig.~\ref{f:gluonTMDs}, we show the unpolarised TMD gluon PDF as a function of the intrinsic transverse momentum $|\boldsymbol{k}_\perp|$. In the left panel, we show the TMD PDFs at the nominal scales $\mu = \sqrt{\zeta} = M_H$ for different values of $x$, namely $x=0.001, 0.01, 0.1$. In the right panel, we show the TMD PDF at $x=0.01$ at three different values of $Q=\mu = \sqrt{\zeta}$, specifically $Q=6$~GeV, which roughly corresponds to the production of a pair of $J/\psi$'s, the usual $M_H$, and $Q=350$~GeV, which is the typical energy for the production of a $t\overline{t}$ pair. Bands correspond to one-sigma uncertainties. 

We observe the gluon TMD grows in size as $x$ decreases. In fact, its $|\boldsymbol{k}_\perp|$-integral is related to the gluon collinear PDFs, which indeed grows very rapidly at large scales and small values of $x$. We also stress that the behaviour in $x$ of the gluon TMD is fully driven by the perturbative matching and collinear PDFs because our model for $f_{\rm NP}$ is independent of $x$.
Finally, we observe that the gluon TMD becomes broader at larger values of $Q$ as a consequence of the evolution.

\subsection{Perturbative convergence}
\label{ss:pertconv}

In previous sections we discussed our baseline fit, which was obtained at N$^3$LL accuracy. We now show the importance of including higher-order perturbative corrections in the theoretical predictions to achieve a good description of the experimental data. To this end, we perform additional fits at lower perturbative accuracy, namely at NLL', NNLL, and NNLL'~\cite{Bacchetta:2019sam}, and compare the results with the baseline N$^3$LL fit. We do not consider LL and NLL accuracies because in both cases the data/theory agreement is very poor. For these additional fits, we use the baseline cut $q_T/M_H<0.3$.

\begin{figure}[h]
\centering
\includegraphics[width=0.49\linewidth]{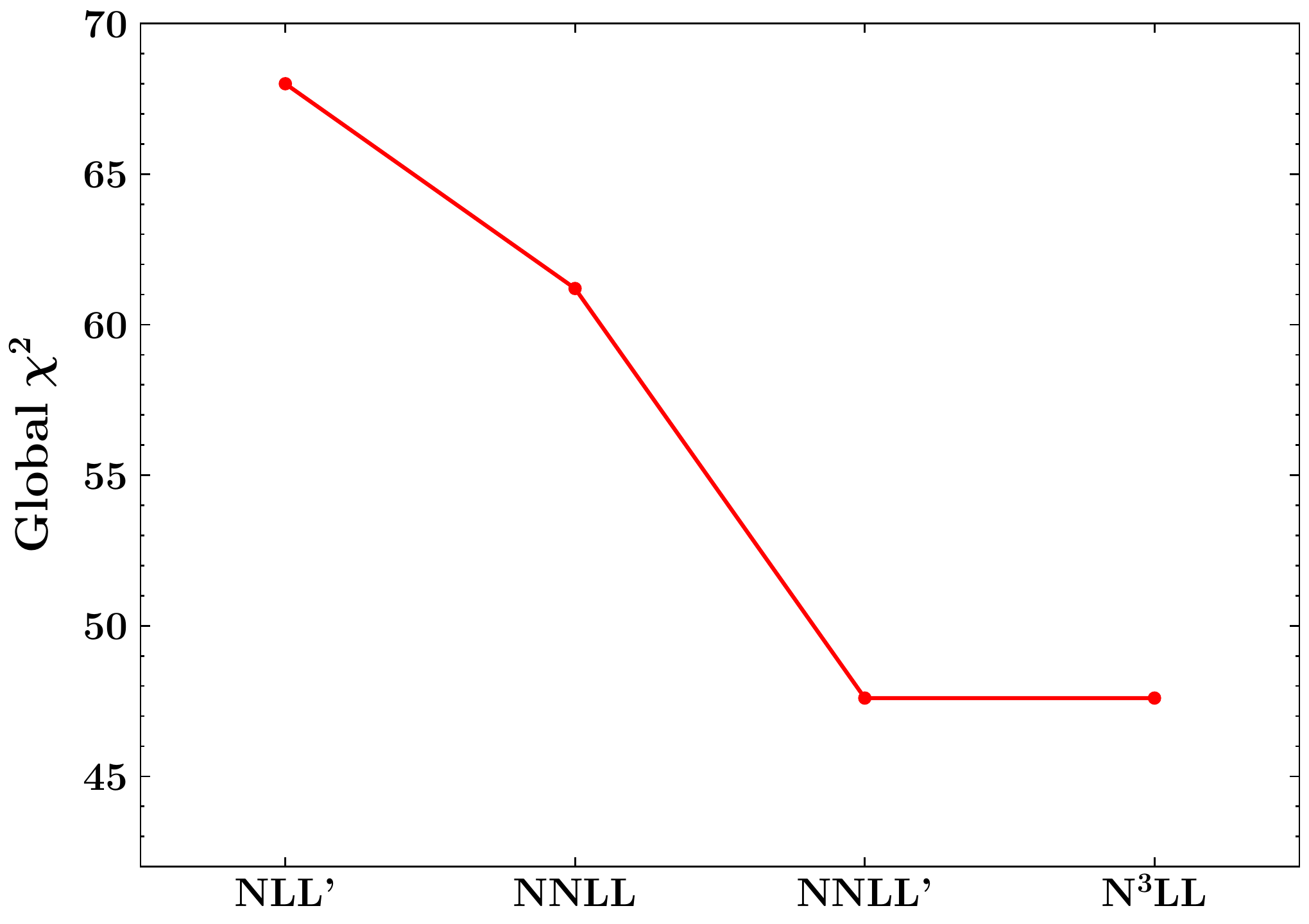}
\caption{Values of the global $\chi^2$ of the fits at NLL', NNLL, NNLL' and N$^3$LL perturbative accuracies.}
\label{f:pertacc_chi2}
\end{figure}

In Fig.~\ref{f:pertacc_chi2}, we show the behaviour of the global $\chi^2$ as the perturbative accuracy increases. Note that we didn't divide for the number $N_{\rm dat}$ of data points in order to better appreciate the size of the differences.

We observe that the agreement between data and theory improves significantly moving from NLL' to NNLL', while it remains essentially unchanged taking a further step to N$^3$LL. This indicates that the perturbative series is converging and that the inclusion of higher-order perturbative corrections is important to achieve an accurate description of data. We conclude that, given the current experimental uncertainties, NNLL' accuracy or higher is appropriate to reliably extract gluon TMDs.

\begin{figure}[h]
\centering
\includegraphics[width=0.65\linewidth]{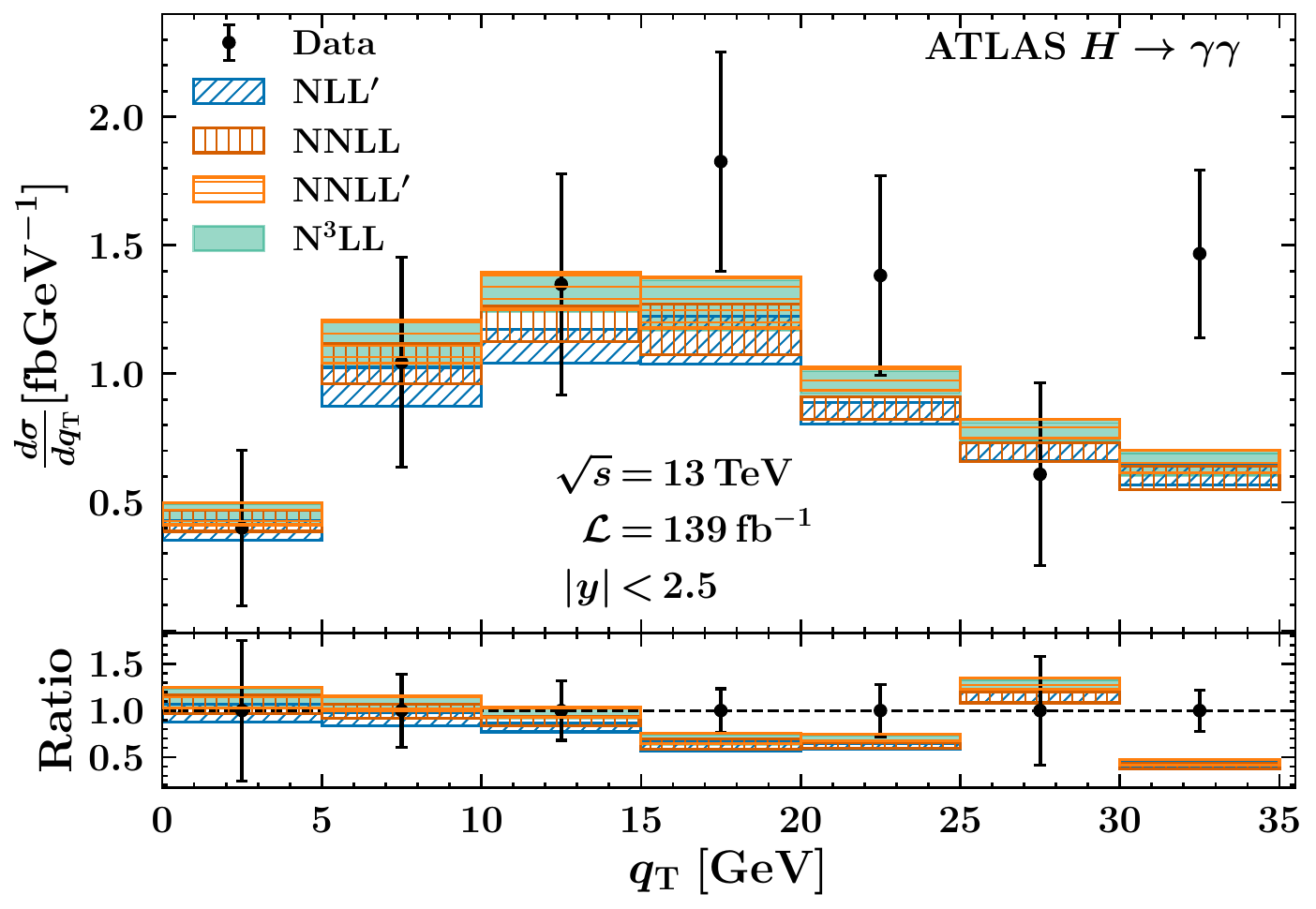}
\caption{Comparison between experimental data for the ATLAS Run II measurements at 13 TeV and the theoretical predictions obtained from the fits at NLL', NNLL, NNLL', and N$^3$LL. The layout of the plot is the same as in Fig.~\ref{f:DataTheoComp}.}
\label{f:pertacc_obs}
\end{figure}

In order to better visualise the impact of higher-order correction, in Fig.~\ref{f:pertacc_obs} we compare fitted theoretical predictions for all of the available perturbative orders to the ATLAS Run II ($H\to \gamma\gamma$) dataset.
We see that the agreement between data and theory improves as the perturbative accuracy increases, with the NNLL' and N$^3$LL predictions being able to reproduce at best both shape and normalisation of the dataset. In particular, the inclusion of higher-order corrections helps describe the peak and the tail of the distribution at relatively large values of $q_T$, which is the region where most of the tension between data and theory is observed at lower perturbative orders.

\subsection{Dependence on the $q_T$ cut}
\label{ss:qTcut}

Finally, in this section we discuss the dependence on the cut on the measured transverse momentum $q_T$. This is a key ingredient of any TMD fit since it restricts the analysis to the region where TMD factorisation is valid. In our baseline analysis, we include all data points with $q_T / M_H < 0.3$. Given the scarcity of Higgs data, this choice has the scope of retaining as much experimental information as possible while staying in the applicability region of TMD factorisation. Nonetheless, it is instructive to explore the effect of imposing more conservative cuts. In Tab.~\ref{t:chi2_vs_qToQ}, we report the values of the global $\chi^2 / N_{\rm dat}$ of the N$^3$LL fit with the corresponding extracted value of the parameter $g_1$ for three different choice of the cut, namely $q_T / M_H < 0.2$, $0.25$, and the baseline cut $0.3$.
\begin{table}[h] \centering \setlength{\tabcolsep}{20pt} \renewcommand{\arraystretch}{1.15} \begin{tabular}{lccc} \hline
                                                                                              Cut                      & $N_{\rm dat}$ & $\chi^2 / N_{\rm dat}$ & $g_1 \big[ \text{GeV}^2 \big]$ \\
                                                                                              \hline
                                                                                              $q_T / M_H < 0.2$          & 20            & 1.08                   & $ 7.1 \pm 4.3$  \\
                                                                                              $q_T / M_H < 0.25$         & 28            & 1.35                   & $12.6 \pm 4.7$  \\
                                                                                              \bm{$q_T / M_H < 0.3$} & \textbf{30}   & \textbf{1.49}              & \bm{$14.4 \pm 5.1$}  \\
                                                                                              \hline \end{tabular} \caption{Values of the global $\chi^2 / N_{\rm dat}$ and extracted $g_1$ parameter of the fits at N$^3$LL for different choices of the cut on $q_T / M_H$. The number of data points $N_{\rm dat}$ included in each fit is also reported.}
\label{t:chi2_vs_qToQ}
\end{table}
We observe that the data/theory agreement increases for more conservative cuts. In particular, the global $\chi^2 / N_{\rm dat}$ decreases from 1.49 for the baseline cut to 1.08 for the most conservative cut. This is an indication that some of the tension observed in the fit with the baseline cut can be attributed to data points in the region of relatively large $q_T$. However, the price to pay is a significant reduction of the number of data points, from 30 to 20, which also hampers a sound statistical interpretation of the results.

\begin{figure}[h]
\centering
\includegraphics[width=0.49\linewidth]{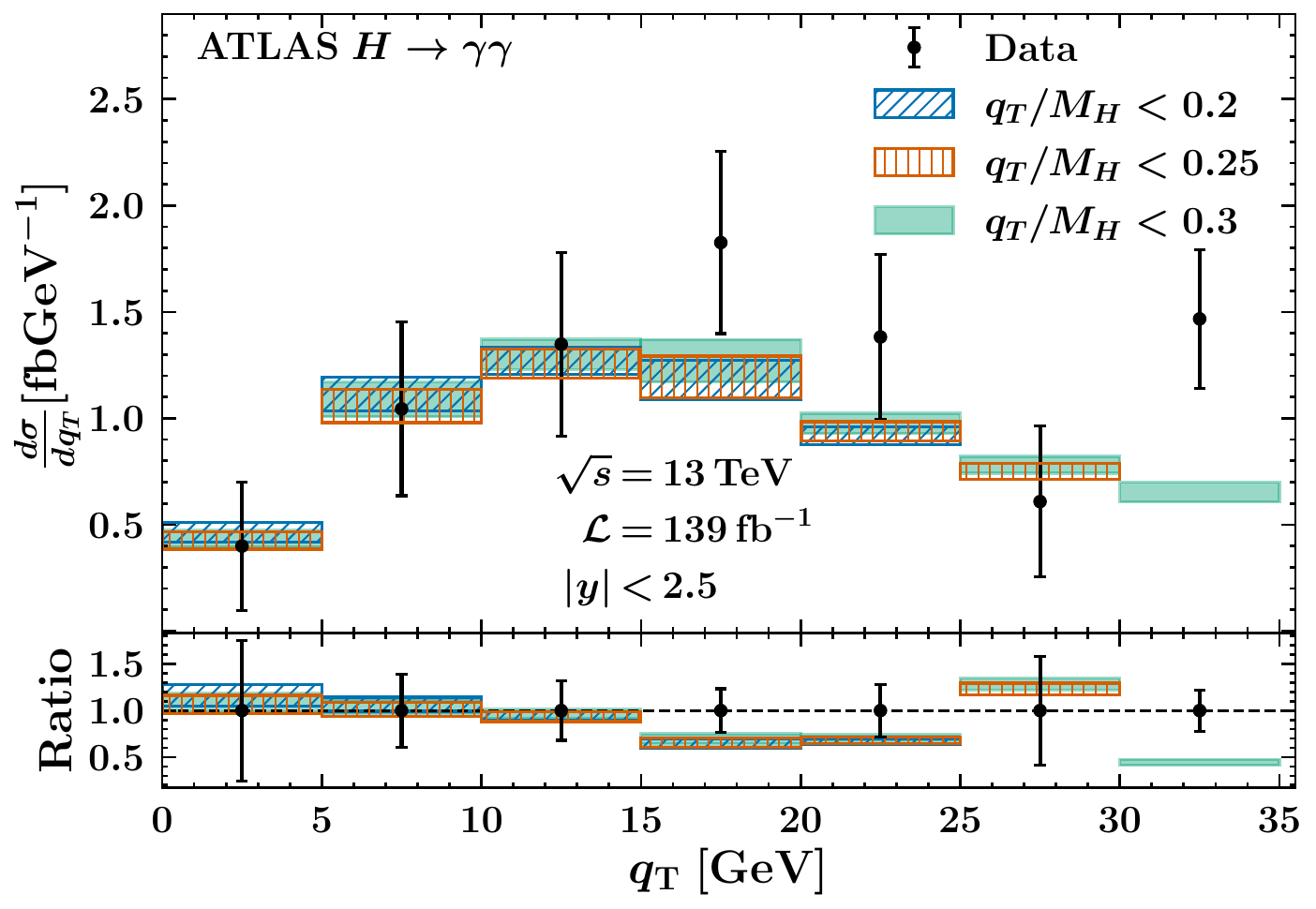}
\includegraphics[width=0.49\linewidth]{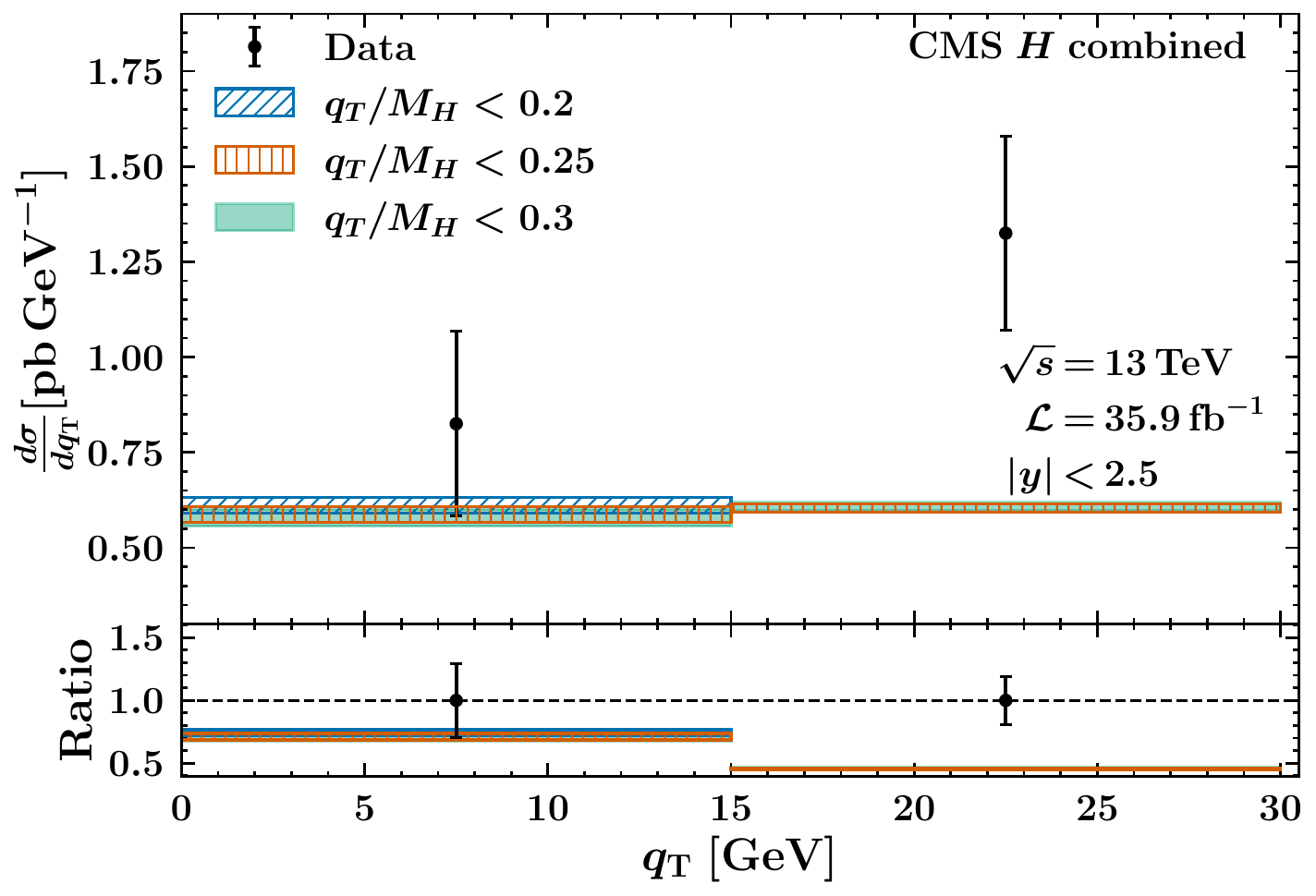}
\caption{Comparison between experimental data for the ATLAS Run II (left panel) and CMS Run II combined (right panel) measurements at 13 TeV and the theoretical predictions obtained from the fits with different cuts on $q_T / M_H$. The layout of the plot is the same as in Fig.~\ref{f:DataTheoComp}.}
\label{f:qToQcut_obs}
\end{figure}

In order to better understand the origin of this tension, in Fig.~\ref{f:qToQcut_obs} we compare the fitted theoretical predictions for the three different cuts to the ATLAS Run II ($H\to \gamma\gamma$) and CMS Run II combined datasets.
We observe that the tension between data and theory is mostly driven by statistical fluctuations in the tail of the distribution at relatively large $q_T$. The tension is particularly evident for the CMS Run II combined dataset, which is the one with the largest $\chi^2 / N_{\rm dat}$ in Tab.~\ref{tab:chi2}. Indeed, the $\chi^2 / N_{\rm dat}$ for this dataset decreases from 4.46 for the baseline cut to 0.76 for the most conservative cut.  For the data points included in all three cases, the agreement between data and theory is similar, which indicates that our choice of the baseline cut is not significantly biasing the TMD.

In the last column of Tab.~\ref{t:chi2_vs_qToQ}, we report the value of the extracted $g_1$ parameter for the three different cuts. We observe that the value of $g_1$ decreases for more conservative cuts, which is an indication that the data points in the region of relatively large $q_T$ are driving the fit towards larger values of $g_1$. However, given the current experimental uncertainties, the values of $g_1$ obtained with different cuts are compatible within one-sigma error. This is yet another indication that the choice of the baseline cut is not significantly affecting the fit results.

\begin{figure}[h]
\centering
\includegraphics[width=0.49\linewidth]{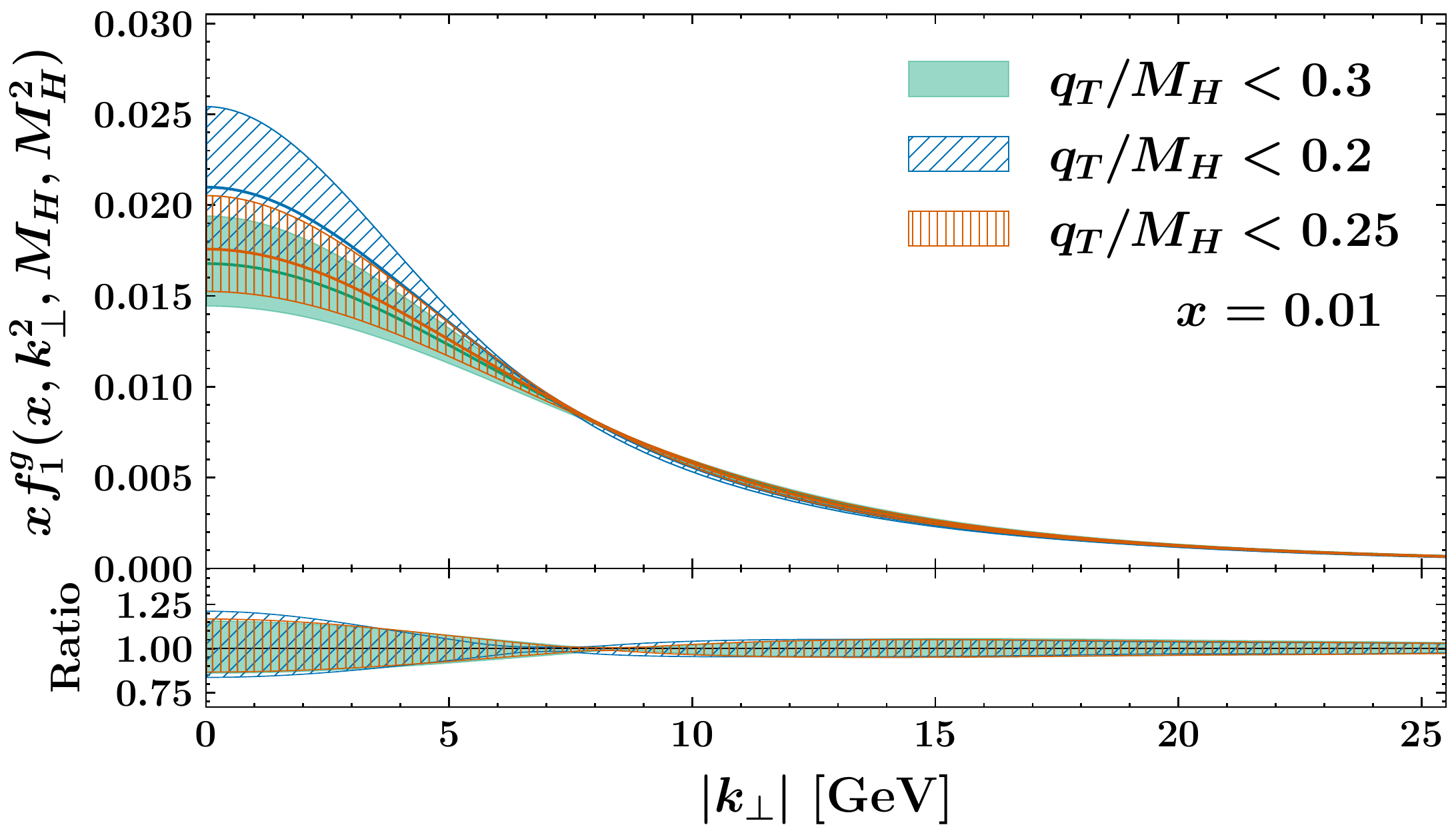}
\caption{The TMD PDF of the gluon in a proton at $\mu = \sqrt{\zeta} = M_H$ and $x=0.01$ as a function of the partonic transverse momentum $|\boldsymbol{k}_\perp|$ obtained from the fits with different cuts on $q_T / M_H$. Uncertainty bands correspond to one-sigma error.}
\label{f:gluonTMDs_qToQ}
\end{figure}

In order to better visualise the effect of the different cuts, in Fig.~\ref{f:gluonTMDs_qToQ} we compare the gluon TMD at $\mu = \sqrt{\zeta}=M_H$ and $x=0.01$ obtained from the fits with different cuts on $q_T$.
As expected, distributions obtained with different cuts are compatible within one-sigma error. We also see that the TMD obtained with the most conservative cut is slightly less broad than the one obtained with the baseline cut, which is consequence of the decrease of the $g_1$. This is an indication that the data points in the region of relatively large $q_T$ are driving the fit towards smaller TMDs at low $|\boldsymbol{k}_\perp|$. However, given the current experimental uncertainties, this effect is not statistically significant.

\section{Conclusion}
\label{s:conclusion}

In this work, we have presented the first extraction of the
unpolarised gluon TMD $f_1^g$ from LHC measurements of the Higgs-boson
transverse-momentum distribution. The analysis is based on the
complete set of available ATLAS and CMS data at $\sqrt{s}=8$ and $13$
TeV in the diphoton and four-lepton decay channels, and is carried out
within the TMD factorisation framework at N$^3$LL accuracy, including
the linearly polarised gluon TMD $h_1^{\perp g}$ with its NNLO
matching coefficient. The fit is implemented in the NangaParbat
framework of the MAP Collaboration, and fiducial selections are
consistently incorporated through a phase-space reduction factor
computed for the relevant decay topologies.

The baseline fit, performed with the cut $q_T/M_H < 0.3$, reproduces
both the shape and the normalisation of the experimental data with a
global $\chi^2/N_{\rm dat} = 1.49$, and yields an intrinsic
nonperturbative parameter $g_1 = 14.4 \pm 5.1\,\text{GeV}^2$. The
relative uncertainty of about $35\%$ reflects the limited sensitivity
of current Higgs data to the nonperturbative content of gluon TMDs, a
direct consequence of the large hard scale $M_H$ at which the
perturbative contribution dominates. The narrow-width approximation,
together with the single scale spanned by the dataset, prevents us
from separately constraining the nonperturbative component of the
Collins-Soper kernel; the latter has therefore been estimated by
rescaling the corresponding quark coefficient extracted in
Ref.~\cite{Avkhadiev:2025wps} by $C_A/C_F$. The study of perturbative
convergence shows a marked improvement of the data/theory agreement
from NLL$'$ to NNLL$'$, and only mild changes when moving from NNLL$'$
to N$^3$LL, indicating that the perturbative series has effectively
stabilised at the highest accuracy considered. Variations of the $q_T$
cut have a moderate impact on the extracted distribution: all values
of $g_1$ obtained with different cuts are mutually compatible within
one-sigma uncertainties, suggesting that the baseline choice does not
introduce a significant bias in the extracted TMD.

Several directions naturally follow from this analysis. First, the
inclusion of gluon-sensitive observables, possibly measured at future
fixed-target experiments~\cite{Hadjidakis:2018ifr,Aidala:2019pit}, at
scales significantly different from $M_H$. One such example is
$J/\psi$-pair or other quarkonium production, which would enable a
direct determination of the nonperturbative Collins-Soper kernel for
gluons and disentangle the intrinsic component from the evolution
contribution. Second, the increased statistical precision expected
from Run III and the high-luminosity LHC programme should
substantially tighten the determination of the nonperturbative
parameters and eventually provide access to the $x$ dependence of
gluon TMDs, which is presently driven entirely by the perturbative
matching onto collinear PDFs. Third, more flexible parametrisations of
$f_{\rm NP}$, possibly distinguishing between $f_1^g$ and
$h_1^{\perp g}$ and incorporating an explicit $x$ dependence, will
become meaningful as the data sample grows. We leave these
developments to future work.

\begin{acknowledgments}
  We would like to thank C. Pisano for helpful suggestions regarding
  the manuscript, N. Berger and P. Bortignon for insights on fiducial
  acceptance factors in ATLAS and CMS data.  Initial stages of this
  work were performed during the ``Hadron Physics 2030'' program at
  the Institut Pascal (Université Paris-Saclay) with the support of
  the program ``Investissements d'avenir'' ANR-11-IDEX-0003-01.  The
  work of S.A. and G.B. is supported by Fondazione di Sardegna through
  the project ``Journey to the center of the proton'',
  No. F23C25000150007. The work of V.B. has been supported by l'Agence
  Nationale de la Recherche (ANR), project ANR-24-CE31-7061-01.
\end{acknowledgments}

\appendix

\section{Phase-space reduction factor}\label{sec:PSRED}

In this appendix we sketch the computation of the phase-space
reduction factor $\mathcal{P}$ introduced in
Eq.~\eqref{eq:higgsfactoriz}, which encodes the effect of the
kinematic cuts applied to the particles in the final state.  The
factor $\mathcal{P}$ accounts for the impact of fiducial selections on
the theoretical cross section. It is defined, for a generic
$N$-particle final state, as the ratio of the squared decay amplitude
integrated over the fiducial region to the same quantity integrated
over the full phase space, namely
\begin{equation}\label{eq:PSredGen}
\mathcal{P}(q) \;=\; \frac{\displaystyle \int_{\text{fid.\,reg.}}\!d\Phi_N\;\bigl|\mathcal{M}(p_1,\ldots,p_N)\bigr|^{2}}
{\displaystyle \int d\Phi_N\;\bigl|\mathcal{M}(p_1,\ldots,p_N)\bigr|^{2}}\,,
\end{equation}
where $q$ is the four-momentum of the decaying particle,
$\mathcal{M}$ is the amplitude of the decay process and $d\Phi_N$ is
the $N$-body Lorentz-invariant phase space,\footnote{Throughout this
appendix we neglect the mass of the final-state particles, so that
the on-shell condition reduces to $p_i^{2}=0$.}
\begin{equation}\label{eq:PhiN}
d\Phi_N(q;p_1,\ldots,p_N) \;=\; (2\pi)^{4}\,\delta^{(4)}\!\left(q-\sum_{i=1}^{N}p_i\right)\prod_{i=1}^{N}\frac{d^{4}p_i}{(2\pi)^{3}}\,\delta(p_i^{2})\,\theta(p_{i,0})\,.
\end{equation}
In the following, we discuss the computation of $\mathcal{P}$ for the
two-body and four-body final states relevant to the diphoton and
four-lepton decay channels of the Higgs boson, respectively.  We first
treat the diphoton channel, $H\to\gamma\gamma$ and then we extend the
construction to the four-lepton channel, $H\to 4\ell$.

\subsection{Cuts on the two-particle final state}\label{app:P2g}

For the diphoton final state, Eq.~\eqref{eq:PSredGen} reads
\begin{equation}\label{eq:PSredDef}
\mathcal{P}(q) =\frac{\displaystyle \int_{\text{fid.\,reg.}}\!d^{4}p_1\,d^{4}p_2\;\delta(p_1^{2})\,\delta(p_2^{2})\,\theta(p_{1,0})\,\theta(p_{2,0})\,\delta^{(4)}(p_1+p_2-q)\,\bigl|\mathcal{M}(p_1,p_2)\bigr|^{2}}
{\displaystyle \int d^{4}p_1\,d^{4}p_2\;\delta(p_1^{2})\,\delta(p_2^{2})\,\theta(p_{1,0})\,\theta(p_{2,0})\,\delta^{(4)}(p_1+p_2-q)\,\bigl|\mathcal{M}(p_1,p_2)\bigr|^{2}}\,,
\end{equation}
where $p_1$ and $p_2$ are the four-momenta of the outgoing photons,
$q=p_1+p_2$, and $\mathcal{M}(p_1,p_2)$ is the amplitude for the decay
$H\to\gamma\gamma$.  This latter can be written as
\begin{equation}\label{eq:Mgg}
\mathcal{M} \;=\; \frac{e^{2}\,g}{(4\pi)^{2}\,M_W}\,F\,
\bigl(p_1\!\cdot\!p_2\,g^{\mu\nu}-p_2^{\mu}\,p_1^{\nu}\bigr)\,\epsilon_{\mu}(p_1)\,\epsilon_{\nu}(p_2)\,,
\end{equation}
where $e$ is the electromagnetic coupling, $g$ is the weak coupling,
$M_W$ is the $W$-boson mass and $\epsilon_{\mu}(p_i)$ is the photon
polarisation vector.  The function $F$ collects the contributions of
$W$ and fermion loops,
\begin{equation}
F \;=\; F_W(\beta_W) \;+\; \sum_{f} N_c\,Q_f^{\,2}\,F_f(\beta_f)\,,
\end{equation}
with $N_c=1$ for leptons and $N_c=3$ for quarks, and
\begin{equation}
\beta_W \;=\; \frac{4 M_W^{2}}{M_H^{2}}\,,\qquad
\beta_f \;=\; \frac{4 m_f^{2}}{M_H^{2}}\,,
\end{equation}
while
\begin{equation}
\begin{aligned}
F_W(\beta) &\;=\; 2 + 3\beta + 3\beta(2-\beta)\,f(\beta)\,,\\
F_f(\beta) &\;=\; -2\beta\,\bigl[\,1+(1-\beta)\,f(\beta)\,\bigr]\,,
\end{aligned}
\end{equation}
with $f(\beta)$ the function defined in Eq.~\eqref{eq:fdecayfunction}.
Summing over photon polarisations, one obtains
\begin{equation}\label{eq:Mgg2}
\bigl|\mathcal{M}\bigr|^{2} \;\propto\; F^{2}\,\frac{(p_1+p_2)^{4}}{2}\,,
\end{equation}
so that Eq.~\eqref{eq:PSredDef} becomes
\begin{equation}\label{eq:PSredDef2}
\mathcal{P}(q) \;=\;
\frac{\displaystyle \int_{\text{fid.\,reg.}}\!d^{4}p_1\,d^{4}p_2\,\delta(p_1^{2})\,\delta(p_2^{2})\,\theta(p_{1,0})\,\theta(p_{2,0})\,\delta^{(4)}(p_1+p_2-q)\,(p_1+p_2)^{4}}
{\displaystyle \int d^{4}p_1\,d^{4}p_2\,\delta(p_1^{2})\,\delta(p_2^{2})\,\theta(p_{1,0})\,\theta(p_{2,0})\,\delta^{(4)}(p_1+p_2-q)\,(p_1+p_2)^{4}}\,.
\end{equation}
The restriction of the integration domain to the fiducial region can be
implemented through a generalised step function $\Theta(p_1,p_2)$,
which equals one inside the fiducial region and vanishes outside.
This allows one to extend the numerator integration to the full
two-photon phase space.  Exploiting momentum conservation through the
four-dimensional $\delta$-function both in the numerator and in the
denominator, and dropping the overall factor $(p_1+p_2)^{4}=q^{4}$,
which is independent of the integration variables, the phase-space reduction factor takes the compact form
\begin{equation}\label{eq:PSredDef4}
\mathcal{P}(q) \;=\;
\frac{\displaystyle \int d^{4}p\,\delta(p^{2})\,\delta\bigl((q-p)^{2}\bigr)\,\theta(p_{0})\,\theta(q_0-p_{0})\,\Theta(p,q-p)}
{\displaystyle \int d^{4}p\,\delta(p^{2})\,\delta\bigl((q-p)^{2}\bigr)\,\theta(p_{0})\,\theta(q_0-p_{0})}\,.
\end{equation}
The two outgoing photons are subject to asymmetric, hierarchy-dependent
selections, as summarised in Table~\ref{tab:photoncuts}.  Following the
standard experimental convention, the leading (subleading) photon is
defined as the one with the largest (smallest) transverse momentum, and
a different $p_T$ threshold is applied accordingly.  Because of asymmetric cuts on the photon rapidities, see Table~\ref{tab:photoncuts}, an analytic evaluation of
Eq.~\eqref{eq:PSredDef4} is not feasible, and we rely on a Monte Carlo integration.

\subsection{Cuts on the four-particle final state}\label{app:P4l}

We now extend the calculation of the phase-space reduction factor
$\mathcal{P}$ to the four-lepton final state,
\begin{equation}\label{eq:psrf4l}
  \mathcal{P}(q) \;=\;
  \frac{\displaystyle \int_{\text{fid.\,reg.}}\!d\Phi_4\;\bigl|\mathcal{M}(p_1,p_2,p_3,p_4)\bigr|^{2}}
  {\displaystyle \int d\Phi_4\;\bigl|\mathcal{M}(p_1,p_2,p_3,p_4)\bigr|^{2}}\,,
\end{equation}
where $p_i$ ($i=1,\ldots,4$) are the lepton four-momenta,
$\mathcal{M}$ is the amplitude for the decay $H\to ZZ\to 4\ell$, and
$d\Phi_4$ is the four-body Lorentz-invariant phase space defined
according to Eq.~\eqref{eq:PhiN}. The decay amplitude can be written
in terms of the $HZZ$ vertex, the two $Z$ propagators, and the leptonic
currents.  Recalling the $Z$-boson couplings to fermions,
\begin{equation}
g_Z \;=\; \frac{e}{\sin\theta_W\,\cos\theta_W}\,,\qquad
v_f \;=\; I_{3,f}-2\,Q_f\,\sin^{2}\theta_W\,,\qquad
a_f \;=\; I_{3,f}\,,
\end{equation}
with $\theta_W$ the weak mixing angle, $I_{3,f}$ the third component
of the weak isospin of the fermion $f$, and $Q_f$ its electric charge
(for charged leptons $I_{3,\ell}=-\tfrac12$ and $Q_\ell=-1$), the
amplitude reads
\begin{equation}
\begin{aligned}
\mathcal{M} \;=\; i\,g_Z\,M_Z\,g_{\mu\nu}\,
&\Biggl\{\frac{-i\,g^{\mu\lambda}}{(p_1+p_2)^{2}-M_Z^{2}+i\,\Gamma_Z M_Z}\;
\frac{-i\,g_Z}{2}\,\bar{u}^{\,s_1}_{e}(p_1)\,\gamma_{\lambda}(v_\ell-a_\ell\gamma_5)\,v^{\,s_2}_{e}(p_2)\Biggr\}\\[2pt]
\times\;
& \Biggl\{\frac{-i\,g^{\nu\rho}}{(p_3+p_4)^{2}-M_Z^{2}+i\,\Gamma_Z M_Z}\;
\frac{-i\,g_Z}{2}\,\bar{u}^{\,s_3}_{\mu}(p_3)\,\gamma_{\rho}(v_\ell-a_\ell\gamma_5)\,v^{\,s_4}_{\mu}(p_4)\Biggr\}\,,
\end{aligned}
\end{equation}
where $M_Z$ and $\Gamma_Z$ are mass and width of the $Z$ boson, and
$s_i$ are the lepton spin indices.  After summing over the lepton
spins, the squared amplitude takes the form
\begin{equation}\label{eq:M4l2}
\bigl|\mathcal{M}\bigr|^{2}\;\propto\;
\frac{\bigl[(v_\ell^{2}+a_\ell^{2})^{2}+4\,v_\ell^{2}\,a_\ell^{2}\bigr]\,(p_1\!\cdot\!p_3)(p_2\!\cdot\!p_4)
\;+\;(v_\ell^{2}-a_\ell^{2})^{2}\,(p_1\!\cdot\!p_4)(p_2\!\cdot\!p_3)}
{\bigl[\bigl((p_1+p_2)^{2}-M_Z^{2}\bigr)^{2}+\Gamma_Z^{2}M_Z^{2}\bigr]\,
 \bigl[\bigl((p_3+p_4)^{2}-M_Z^{2}\bigr)^{2}+\Gamma_Z^{2}M_Z^{2}\bigr]}\,.
\end{equation}
The explicit dependence on the individual lepton momenta in
Eq.~\eqref{eq:M4l2} prevents any further analytic reduction of
Eq.~\eqref{eq:psrf4l}, and a
numerical evaluation is required.

To enable an efficient numerical evaluation and a transparent
implementation of the fiducial cuts, we factorise the four-body
phase space into a sequence of nested two-body phase spaces, in
accordance with the tree-level topology of the decay
$H\to ZZ\to 4\ell$.  We introduce the intermediate four-momenta
\begin{equation}
q_1 \;=\; p_1+p_2\,,\qquad q_2 \;=\; p_3+p_4\,,\qquad q_1+q_2 \;=\; q\,,
\end{equation}
together with their virtualities $Q_1^{2}\equiv q_1^{2}$ and
$Q_2^{2}\equiv q_2^{2}$.  Inserting two unit factors of the form
$\int dQ_i^{2}\,\delta(q_i^{2}-Q_i^{2})$ into Eq.~\eqref{eq:PhiN}, the
four-body phase space factorises as
\begin{equation}\label{eq:PS4_factor}
d\Phi_{4}(q;p_1,p_2,p_3,p_4) \;=\;
\frac{dQ_1^{2}\,dQ_2^{2}}{(2\pi)^{2}}\;
d\Phi_{2}(q;q_1,q_2)\;
d\Phi_{2}(q_1;p_1,p_2)\;
d\Phi_{2}(q_2;p_3,p_4)\,.
\end{equation}
This decomposition is particularly well suited for a Monte Carlo
implementation: each two-body sub-phase space is parametrised by two
angular variables in the corresponding rest frame, while the
virtualities $Q_1^{2}$ and $Q_2^{2}$ are sampled independently with an
importance-sampling weight tuned on the two $Z$ Breit-Wigner peaks of
Eq.~\eqref{eq:M4l2}.  This procedure reduces the effective
dimensionality of the integration while exposing simple two-body
kinematics, and allows the fiducial selections of
Table~\ref{tab:leptoncuts} ($p_T$ thresholds, $|\eta|$ windows,
$m_{Z_1}/m_{Z_2}$ mass windows, ordered-lepton cuts) to be imposed
directly at particle level.  Eq~\eqref{eq:psrf4l} can thus be
evaluated numerically in a stable and efficient way for any value of
the kinematic variables $(Q,y,q_T)$.

\clearpage
\bibliography{references}

\end{document}